\documentclass[final,5p,times,twocolumn,authoryear]{elsarticle}
\usepackage{graphics}
\usepackage{graphicx}
\usepackage{epsfig}
\usepackage{subfigure}
\usepackage{amssymb}

\journal{Journal of Contaminant Hydrology}

\begin{document}

\begin{frontmatter}

\title{Spatial and temporal features of dense contaminant transport: experimental investigation and numerical modeling}

\author{Andrea Zoia\corref{cor1}}
\address{CEA/Saclay, DEN/DM2S/SFME/LSET, B\^at.~454, 91191 Gif-sur-Yvette Cedex, France}
\ead{andrea.zoia@cea.fr}
\cortext[cor1]{Corresponding author}
\author{Christelle Latrille}
\address{CEA/Saclay, DEN/DPC/SECR/L3MR, B\^at.~450, 91191 Gif-sur-Yvette Cedex, France}
\author{Alberto Beccantini}
\address{CEA/Saclay, DEN/DM2S/SFME/LTMF, B\^at.~454, 91191 Gif-sur-Yvette Cedex, France}
\author{Alain Cartalade}
\address{CEA/Saclay, DEN/DM2S/SFME/LSET, B\^at.~454, 91191 Gif-sur-Yvette Cedex, France}

\begin{abstract}
We investigate the spatial and temporal features of dense contaminant plumes dynamics in porous materials. Our analysis is supported by novel experimental results concerning pollutant concentration profiles inside a vertical column setup. We describe the experimental methods and elucidate the salient outcomes of the measurements, with focus on miscible fluids in homogeneous saturated media. By resorting to a finite elements approach, we numerically solve the equations that rule the pollutants migration and compare the simulation results with the experimental data. Finally, we qualitatively explore the interfacial dynamics behavior between the dense contaminant plume and the lighter resident fluid that saturates the column.
\end{abstract}


\end{frontmatter}

\section{Introduction}

The migration of dense pollutant plumes through porous materials is key to mastering such environmental and technological challenges as site remediation, safety assessment for waste repositories and enhanced petroleum recovery~\citep{ophori,yang,holm}, to name a few. Yet, despite being a well-known and long studied problem~\citep{wooding62, bachmat}, it keeps raising many conceptual as well as practical issues~\citep{bear, demarsily, sahimi}; for an overview of recent advances, see, e.g.,~\citep{rev1,rev2}. When contaminants are sufficiently diluted, flow and concentration fields do not affect each other, so that pollutants dynamics can be addressed by first determining the pressure distribution in the traversed region (which imposes velocity within the pore network) and then separately computing the concentration profile due to advection and dispersion mechanisms. Under such conditions, transport is Fickian, i.e., ruled by the standard advection-dispersion equation (ADE)~\citep{bear}, at least for homogeneous media. On the contrary, when contaminant plumes are sufficiently concentrated the pressure field will depend on the fuid density, which in turn is affected by the contaminant concentration. Nonlinear effects come thus into play, by virtue of the coupling between flow and transport (see, e.g.,~\citep{schincariol, gelhar1, gelhar2, wooding}). Based on the density (and in general also viscosity~\citep{jiao}) contrasts between the pollutants-loaded fluid and the resident fluid that permeates the porous material, instabilities (fingerings) may appear at their mutual interface~\citep{manickam, tyler, wooding}. The relevance of such instabilities to the overall contaminant dynamics is determined by the intensity of density and viscosity contrasts~\citep{bacri, jiao, schincariol, gelhar1, dalziel, oltean, rogerson}.

\begin{figure}[t]
\centerline{\epsfxsize=9.0cm\epsfbox{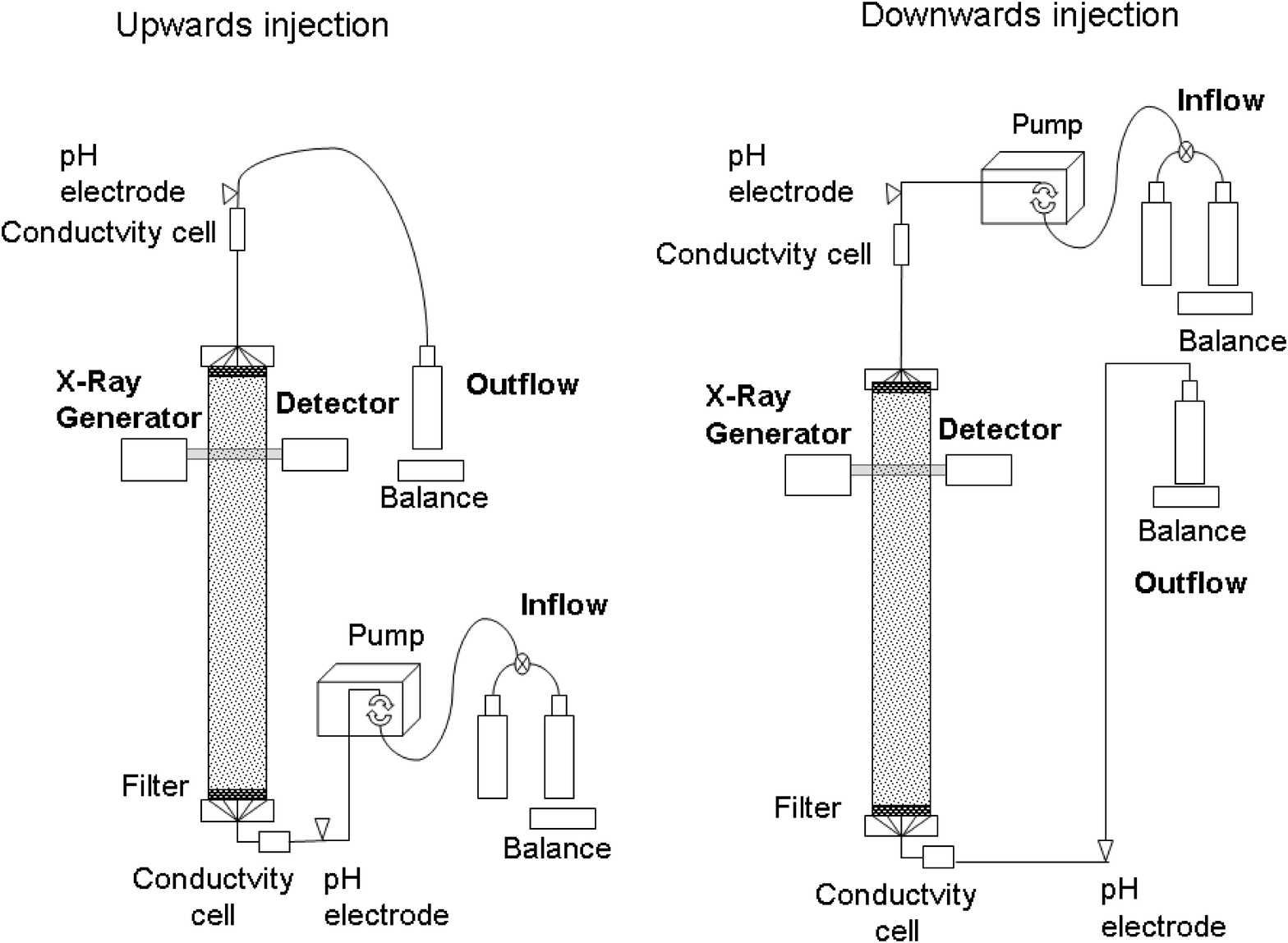}}
\caption{Scheme of the BEETI experimental setup for downwards and upwards tracer injection.}
   \label{fig0}
\end{figure}

High densities and/or density gradients are encountered when either the contaminant itself is strongly concentrated at the source, or the plume flows through regions that are rich in salt (brines)~\citep{hassanizadeh, schincariol, schotting}. In particular, the latter case might become a major concern for radioactive waste disposal near salt domes~\citep{hassanizadeh}. Variable-density migration affects transport by altering both the time scale (average displacement) and the spatial extent (spread) of pollutants plumes~\citep{rev1, rev2}. Therefore, in such conditions the plume dynamics can not be described by resorting to the standard ADE model, which does not account for density effects~\citep{wood}. An exact quantification of the pollutants average displacement and spread depends on correctly capturing the underlying physics of the above mentioned processes: several studies have been performed to this aim, covering homogeneous saturated and heterogenous unsaturated materials, density as well as viscosity contrasts, and different geometries and column configurations; see, e.g.,~\citep{schincariol, wood, jiao, dalziel, oltean, dangelo, simmons, tchelepi, dewit, zimmerman, ruith1, ruith2, debacq, seon, perrin}. A central outcome is that density effects can play a major role in affecting the fate of the contaminant plume even in homogeneous and saturated porous media~\citep{jiao, wood, simmons, oltean}. Moreover, it has been shown that even modest density differences with respect to the reference fluid (which is frequently freshwater) may lead to measurable deviations from the usual Fickian behavior~\citep{schincariol, hassanizadeh}.

The aim of this work is to explore the spatial and temporal features of dense pollutants transport in porous media, with focus on the case of miscible fluids, i.e., fluids having a single phase but different solute concentrations. Our investigation is supported by novel experimental measurements of contaminant profiles inside homogeneous saturated columns, obtained by means of X-ray spectroscopy. As detailed in the following, X-rays measures allow for two complementary informations: on one hand, the contaminant profiles as a function of time, at a given location; on the other, the spatial evolution of such profiles along the column. We are thus able to fully characterize contaminant dynamics by experimental means, which is especially important when the concentration profiles are time-dependent. The effects of unstable interfacial dynamics on the spatial and temporal evolution of contaminant profiles have received so far only limited attention, whereas studies are usually focused on the mixing properties at the interface between two layers of semi-infinite extension (see, e.g.,~\citep{wood, dewit} and references therein). In groundwater contamination, the spill extent is often finite, since the source is limited in space and time~\citep{dewit}; based on this observation, our analysis will focus on finite-duration contaminant injection. In this case, dynamics is transient: stable and unstable fronts (between injected and resident fluid) will in general be present at the same time.

This paper is organized as follows: in Sec.~\ref{experiments} we describe the experimental setup and outline the salient features of the contaminant concentration profiles along the column. Then, we review the underlying physical equations in Sec.~\ref{modeling} and address their numerical integration by resorting to finite elements modeling in Sec.~\ref{numerical}. Simulation results are successively compared to experimental measurements in Sec.~\ref{test}. The qualitative behavior of the interfacial instabilities and the interplay of the different components governing the physical system are explored in Sec.~\ref{qualitative}. Conclusions are finally drawn in Sec.~\ref{conclusions}.

\section{Experimental methods and results}
\label{experiments}

\subsection{The experimental device}

The BEETI experimental setup has been specifically conceived to assess the spatial and temporal dynamics of tracers in relation to the geometry and the physical-chemical conditions of porous media. The system is composed of a vertical polycarbonate column, a controlled hydraulic circuit that enables tracers migration within the column, and a scanning X-ray spectrometer for measurements (Fig.~\ref{fig0}). The column has height $H = 80$ cm and internal radius $r = 2.5$ cm, so that the aspect ratio is $H/2r = 16 \gg 1$. This experimental setup allows for downwards as well as upwards fluid injection, and several kinds of flow regimes and porous materials can be tested, at various saturation and/or heterogeneity conditions.

The scanning X-ray spectrometry system (SXSS) is based on an X-ray generator with a tungsten source. The emerging beam is filtrated by a neodymium window that selects two energy ranges ($20-40$ keV and $50-75$ keV, respectively). After traversing the sand column and its water and tracers content, the beam is measured by a NaI detector. Both source and detector are kept in position by a controlled rack rail that displaces the SXSS along the column. At specified spatial locations, the SXSS waits $60$ seconds (the counting time) and measures the transmitted dichromatic X-rays. The X-ray countings received by the detector depend on the thickness of the traversed layer, the nature of traversed phase (fluid, porous material or tracer) and the counting time. It is possible to discriminate the different components within each column layer by resorting to the Beer-Lambert's law, which is used to convert the transmitted beam intensity to physical quantities (such as bulk density, porosity, or tracer concentration):
\begin{equation}
R=R_0 \exp \left( -\sum_i \mu_i \rho_i x_i \right).
\end{equation}
Here $R_0$ and $R$ denote the number of photons emitted and detected, respectively, per unit time; $i$ is the index of each phase contained in the medium (sand, water, solute, $...$); $\mu_i$ [m$^2$/Kg] is the mass attenuation coefficient, which depends on the phase composition and on the X-ray energy; $\rho_i$ [Kg/m$^{-3}$] is the density, and $x_i$ [m] the thickness of the phase. The parameters $\mu_i$ and $R_0$ are determined by calibrating the SXSS with specific materials specimens, with perfectly measurable thickness and X-ray attenuation very close to that of water or sand. One of the specimens is used as reference for the calibration of $R_0$, so to avoid saturation effects of the NaI detector, due to strong source intensity. Tracer concentration is determined by rewriting the Beer-Lambert's law as a function of the molecular attenuation coefficient, which is linearly proportional to the tracer concentration solution. The X-ray transmitted countings allow quantitatively accessing the tracer concentration inside the column (as a function of time), at a distance $\ell$ from the inlet: we denote this quantity by $C_\ell(t)$. Given the spatial resolution of the X-ray device, $C_\ell(t)$ actually corresponds to a spatially-averaged measure over the transverse direction (i.e., the column section), for any given height $\ell$.

\begin{figure}[t]
\centerline{\epsfxsize=9.0cm\epsfbox{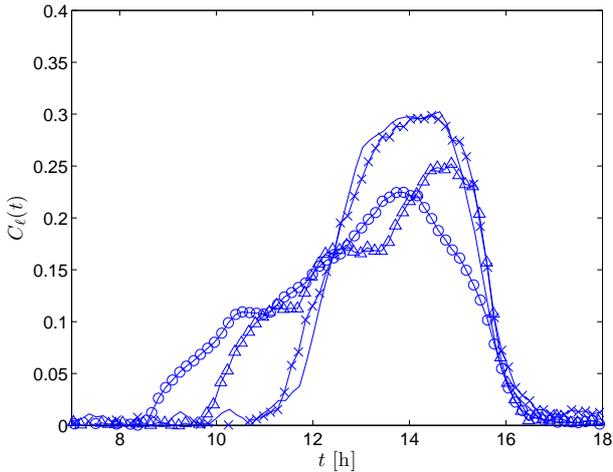}}
\caption{Contaminant concentration curves $C_\ell(t)$ at the last measure point along the column, $\ell=77$ cm, as a function of time [h]. Downwards injection at increasing molarities $C^{mol}=0.05$ (solid line), $0.1$ (crosses), $0.2$ (triangles) and $0.5$ (circles) mol/L.}
   \label{fig1}
\end{figure}

\begin{figure}[t]
\centerline{\epsfxsize=9.0cm\epsfbox{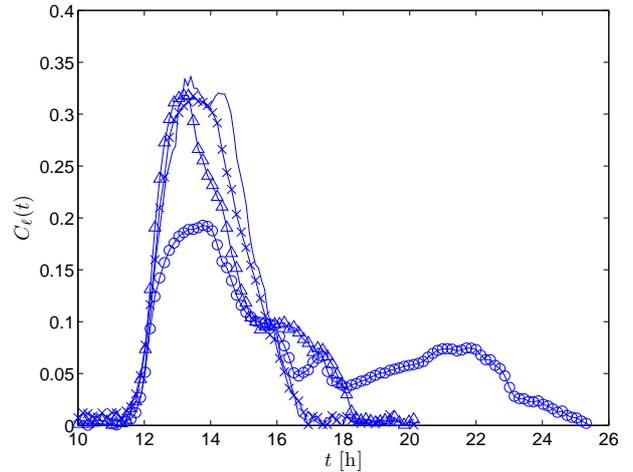}}
\caption{Contaminant concentration curves $C_\ell(t)$ at the last measure point along the column $\ell=77$ cm, as a function of time [h]. Upwards injection at increasing molarities $C^{mol}=0.05$ (solid line), $0.1$ (crosses), $0.2$ (triangles) and $0.5$ (circles) mol/L.}
   \label{fig2}
\end{figure}

In order to obtain a homogeneously packed porous medium, the dry sand was poured continuously and packed while making the column vibrate. Solution saturation was imposed starting from the column bottom. Medium homogeneity and saturation were assessed by dichromatic X-ray spectrometry, which allows estimating bulk density and porosity along the column with a resolution of $5$ mm in the longitudinal direction.

Upwards experiments (Fig.~\ref{fig0}, left) are performed by injecting background and tracer solutions from the lower entrance. Conversely, the fluids are injected from the upper entrance for downwards experiments (Fig.~\ref{fig0}, right). The solutions are drained by a pump which regulates the solution flow rate and the tracer injection at the column inlet. The steady state Darcy flow $q$ is also verified by weighing the outgoing solution. Electric conductivity cells are adopted on-line at the column inlet and outlet, so to monitor tracer displacement through the column. At the outlet of the column, $C_{\ell=H}(t)$ coincides with the breakthrough curve, which is the most frequently measured quantity in contaminant migration experiments and provides information about the tracer mean displacement over the entire column length.

\subsection{Experimental conditions}

\begin{table}[t]
\begin{center}
\begin{tabular}{|c|c|c|}
\hline
Molarity [mol/L] & Density [Kg/m$^{3}$] & $\Delta \rho/\rho_0$ [-] \\
\hline
0.05 & 1004.3 & 0.006\\
\hline
0.1 & 1010.4 & 0.012\\
\hline
0.2 & 1022.4 & 0.024\\
\hline
0.5 & 1058.5 & 0.060\\
\hline
\end{tabular}
\end{center}
\caption{Density variations for KI in water, at increasing solution molarity.}
\label{tab1}
\end{table}

In the following, we refer to a saturated column filled with homogeneously mixed Fontainebleau quartz sand, with bulk density $1.77 \pm 0.01$ g/cm$^{3}$ and average grain diameter $200$ $\mu$m. The sand was previously washed in pure water (MilliQ) and equilibrated with a KCl $10^{-3}$ M solution, so to minimize the ionic exchange between solid sites surface and tracers.

The average porosity is $\theta = 0.333 \pm 0.05$, as measured by the SXSS (to be compared with the value $\theta = 0.328$ as obtained by weighing the sand poured into the column). The dispersivity is $\alpha = 0.1$ cm, as from previous experimental runs (at higher flow rates) on the BEETI device. All measurements are performed at constant room temperature $T = 20^o$C. The reference saturating solution contains KCl at a molar concentration of $C^{mol}=10^{-3}$ mol/L (molar mass equal to $74.5$ g/mol), so that the corresponding reference density is $\rho_0 = 998.3$ Kg/m$^{3}$ at $T = 20^o$C. The injected tracer is KI (molar mass equal to $166$ g/mol), at various molar concentrations. The examined experimental conditions are presented in Tab.~\ref{tab1}, together with the corresponding injected fluid densities $\rho$ and differential densities $\Delta \rho/\rho_0 = (\rho - \rho_0)/\rho_0$, as derived from \citep{otaka}.

The stationary flow is $q = 2$ cm/h and the duration of the flux step injection is $T= 3$ h. In most experimental runs, the measure points are located at a distance $\ell = 7.7$, $23.1$, $38.5$, $46.2$ and $77$ cm from the column inlet.

The experimental conditions are such that clogging or formation of colloidal particles, which could alter the interpretation of the obtained results, can be excluded. Chemical reactions or sorption/desorption phenomena can be ruled out as well. 

\subsection{Experimental results}

\begin{figure}[t]
\centerline{\epsfxsize=9.0cm\epsfbox{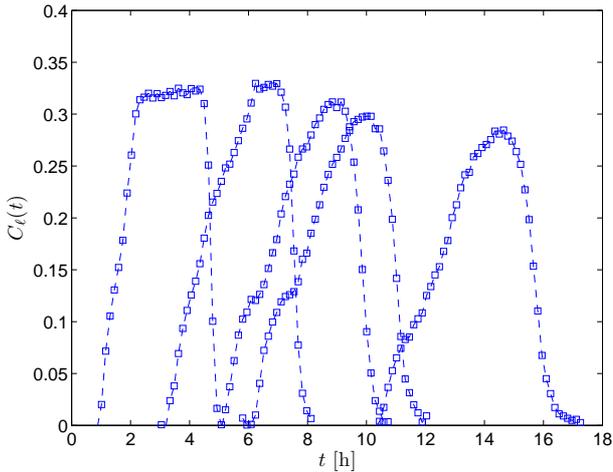}}
\caption{Downwards injection at a reference molarity $C^{mol}=0.2$ mol/L. Contaminant concentration curves $C_\ell(t)$ measured at sections $\ell =7.7, 23.1, 38.5, 46.2,$ and $77$ cm (from left to right), as a function of time [h]. Lines have been added to guide the eye.}
   \label{fig3}
\end{figure}

\begin{figure}[t]
\centerline{\epsfxsize=9.0cm\epsfbox{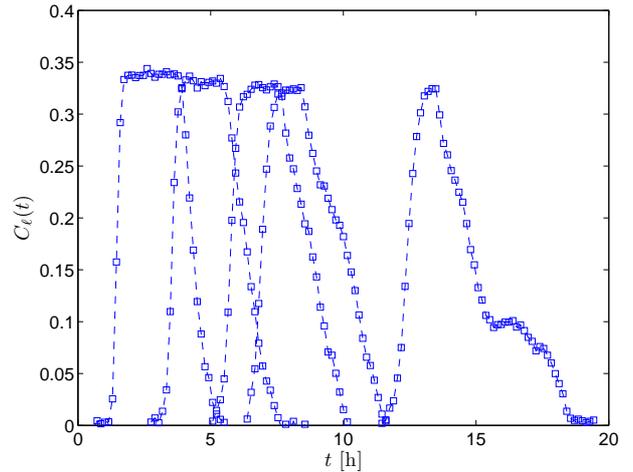}}
\caption{Upwards injection at a reference molarity $C^{mol}=0.2$ mol/L. Contaminant concentration curves $C_\ell(t)$ measured at sections $\ell =7.7, 23.1, 38.5, 46.2,$ and $77$ cm (from left to right), as a function of time [h]. Lines have been added to guide the eye.}
   \label{fig4}
\end{figure}

In the following, we provide some representative results that can best illustrate the set of measures performed with the BEETI device. In Figs.~\ref{fig1} and~\ref{fig2} we start by displaying the concentration profiles corresponding to the last measure point along the column as a function of time, for downwards and upwards injection, respectively. Here and throughout the text, concentration profiles have been normalized to their respective areas. Curves are shown for increasing values of the concentration molarity, ranging from $C^{mol}=0.05$ to $C^{mol}=0.5$ mol/L. At weak molarity, the standard Fickian (symmetrical) profile is recovered. As $C^{mol}$ increases, the curves become increasingly skewed and appreciably deviate from the Gaussian shape. The sign of the skewness depends on the flow direction: downwards injection leads to negatively skewed profiles (i.e., contaminant flows out earlier than expected for a Fickian flux: see Fig.~\ref{fig1}), whereas upwards injection leads to positively skewed profiles (i.e., contaminant flows out later than expected: see Fig.~\ref{fig2}). These effects on transport can be mainly attributed to gravity (via the density coupling). An unstable front appears where a dense fluid overlies a lighter one, and the macroscopic effect is an enhanced diffusivity (i.e., spread) at the interface~\citep{wooding}: this mechanism is known as Rayleigh-Taylor instability~\citep{taylor}. Actually, the interplay between mechanical dispersion due to the structure of the porous medium and the additional dispersion due to interfacial instabilities is key to understanding variable-density transport~\citep{wooding}.

\begin{figure}[t]
\centerline{\epsfxsize=9.0cm\epsfbox{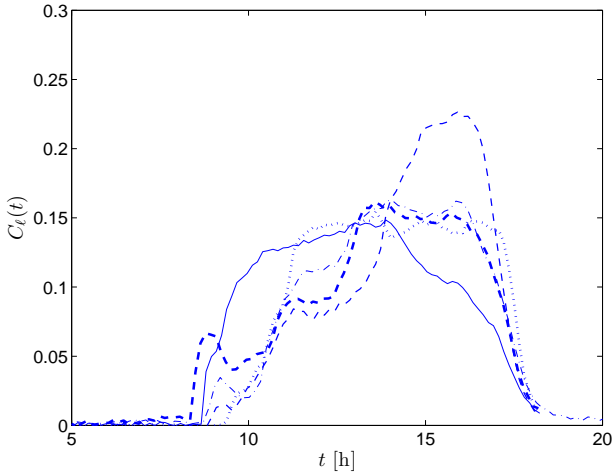}}
\caption{Contaminant concentration curves $C_\ell(t)$ at the last measure point along the column, $\ell=77$ cm, as a function of time [h]. Downwards injection at $C^{mol}=0.5$ mol/L, for $5$ experimental runs.}
   \label{fig5}
\end{figure}

\begin{figure}[t]
\centerline{\epsfxsize=9.0cm\epsfbox{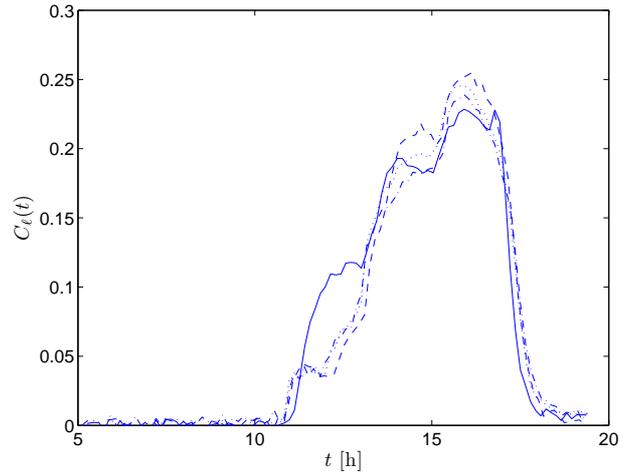}}
\caption{Contaminant concentration curves $C_\ell(t)$ at the last measure point along the column, $\ell=77$ cm, as a function of time [h]. Downwards injection at $C^{mol}=0.2$ mol/L, for $4$ experimental runs.}
   \label{fig6}
\end{figure}

Then, in Figs.~\ref{fig3} and~\ref{fig4} we display the evolution of concentration profiles measured at various heights along the experimental device, for downwards and upwards injection, respectively. In fact, X-rays spectrometry allows accessing contaminant distribution inside the column. Profiles are shown at a fixed molarity $C^{mol}=0.2$ mol/L, which provides a representative case. From these results it appears that the skewness of the curves increases along the column, in the direction dictated by the unstable front. In other words, the portion of the plume that moves faster or slower than the bulk progressively increases. This is because (for the experimental conditions considered here) instabilities can grow, overcoming the smoothing effects of the dispersion induced by the porous material. We remark also that the concentration profiles are almost symmetric upon reversing flow direction, which underlines the prominent role of gravity in determining the pollutants dynamics.

Finally, an important question concerns the reproducibility of the experimental results presented here. As a general remark, the presence of instabilities at the interface between resident and displacing fluids leads to fluctuations and thus to randomness in the shape of contaminant concentration profiles; this holds true even for homogeneous and saturated porous media, such as our column setup~\citep{rev1}. Triggering of unstable fronts and their interplay with stable fronts will be discussed later. At strong molarity (say $C^{mol}=0.5$ mol/L or stronger), the effects of the unstable front dominate, so that a limited reproducibility of the experimental curves is actually observed, because of the aforementioned stochasticity. In Fig.~\ref{fig5} we display results for the repetition of $5$ experimental runs for downwards injection at $C^{mol}=0.5$ mol/L: only a small portion of the curves is reproducible, and fluctuations are evident. Note however that the average displacement and the spread of the contaminant plume seem largely conserved, although the specific shapes of the profiles vary from run to run. 

At weak molarity (say $C^{mol}=0.05$ mol/L or weaker), contaminant profiles are perfectly reproducible within the limits of experimental measures. This is an expected outcome, since in this case the flow and transport mechanisms are almost decoupled and the plume dynamics approaches the standard Fickian behavior. Finally, at intermediate molarity, we experimentally observe a gradual transition between these two regimes. In Fig.~\ref{fig6} we display the repetition of $4$ experimental runs for downwards injection at $C^{mol}=0.2$ mol/L: a good reproducibility of the concentration profiles is found, as the curves are almost superposed, although fluctuations are already visible. Note in particular that superposition is more remarkable on the right portion of the curves, whose shape is determined by the stable front. Comparable results have been obtained also for the case of upwards injection (not shown here).

\section{The coupled flow-transport model}
\label{modeling}

We address now the description of the physical model underlying the observed contaminant dynamics, on the basis of the experimental results exposed above. For sake of simplicity, we commence by representing the actual geometry of the vertical column as a flat rectangular ($2d$) domain $\Omega$, of height $H$ [cm] and base $B$ [cm]: $\Omega = [0, B] \times [0, H]$. We assume that this region is homogeneous and isotropic, as being filled with well-mixed saturated sand. The dynamics of a contaminant plume injected in $\Omega$ is ruled by three equations that condense the physics of the problem, namely fluid mass conservation, advection-dispersion of the contaminant concentration field $c({\bf x},t)$ and Darcy's law for the advection field ${\bf u}({\bf x},t)$. Concerning mass conservation, it is customary to introduce the so-called Boussinesq approximation, i.e., to assume that the effects of density variations in rigid porous media are retained as being significant only when appearing multiplied by the gravity constant $g$ (see, e.g.,~\citep{bear}). This is indeed the case for most examples of dense contaminant transport, where the density variations are of the order of a few percents with respect to the resident fluid~\citep{simmons, rev1, tartakovsky_jfm}. Under this hypothesis, fluid mass balance simplifies to
\begin{equation}
\nabla \cdot{\bf u}({\bf x},t) \simeq 0,
\label{mass}
\end{equation}
\begin{figure}[t]
\centerline{\epsfxsize=9.0cm\epsfbox{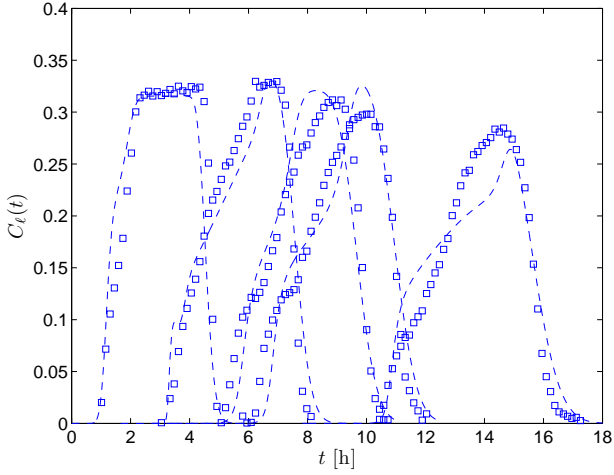}}
\caption{Downwards injection at a reference molarity $C^{mol}=0.2$ mol/L and $q=2$ cm/h. Contaminant concentration curves $C_\ell(t)$ measured at sections $\ell =7.7, 23.1, 38.5, 46.2,$ and $77$ cm (from left to right), as a function of time [h]. Squares represent experimental data, dashed lines model estimates.}
   \label{fig7}
\end{figure}
\begin{figure}[t]
\centerline{\epsfxsize=9.0cm\epsfbox{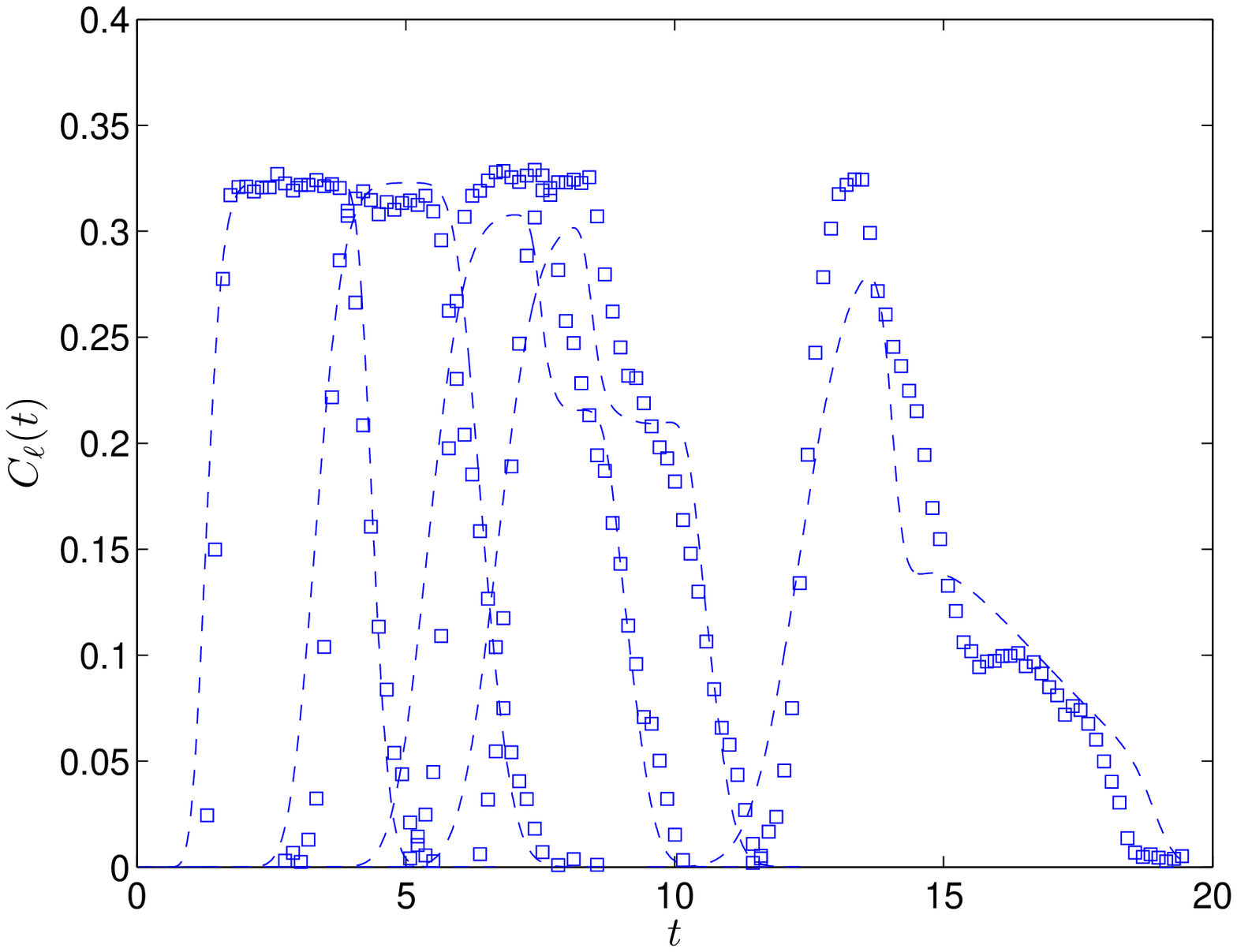}}
\caption{Upwards injection at a reference molarity $C^{mol}=0.2$ mol/L and $q=2$ cm/h. Contaminant concentration curves $C_\ell(t)$ measured at sections $\ell =7.7, 23.1, 38.5, 46.2,$ and $77$ cm (from left to right), as a function of time [h]. Squares represent experimental data, dashed lines model estimates.}
   \label{fig8}
\end{figure}
in the absence of sources and sinks. Mass transport is given by the standard advection-dispersion equation, provided that the traversed material is sufficiently homogeneous~\citep{cortis_homog}:
\begin{equation}
\frac{\partial}{\partial t}\theta c({\bf x},t)=\nabla \cdot \left[\theta D({\bf x},t) \nabla - {\bf u}({\bf x},t) \right]c({\bf x},t),
\label{ade}
\end{equation}
where $\theta$ is the constant average porosity of the medium, ${\bf u}({\bf x},t)$ is the advection field, and the apparent dispersion $D$ includes both mechanical dispersion and molecular diffusion. A widely adopted expression for $D$ reads~\citep{bear}
\begin{equation}
D({\bf x},t)=\frac{\alpha |{\bf u}({\bf x},t)|}{\theta}+ D_0,
\end{equation}
where mechanical dispersion is assumed to be proportional to the absolute value of the pore velocity ${\bf u}/\theta$ through a constant length scale $\alpha$ (the dispersivity) and apparent molecular diffusion has a constant value $D_0$, incorporating the effects of tortuosity. In our experimental conditions, it turns out that $D_0$ is actually negligible with respect to the contribution of mechanical dispersion: $D({\bf x},t) \simeq \alpha |{\bf u}({\bf x},t)|/\theta$. Finally, the advection field is given by the modified Darcy's equation~\citep{bear}
\begin{equation}
{\bf u}({\bf x},t)=-\frac{k}{\mu_c({\bf x},t)}\left[ \nabla p + \rho_c({\bf x},t){\bf g} \right],
\label{darcy}
\end{equation}
where $\nabla p$ is the pressure gradient, ${\bf g}$ is the (vector) gravity constant (with modulus $g$), $k$ is the intrinsic permeability of the porous material (here assumed to be constant, since the medium is homogeneous), $\mu_c({\bf x},t)$ is the fluid viscosity and $\rho_c({\bf x},t)$ is the fluid density. In principle, both density and viscosity depend on concentration. Therefore, in order to close the system of equations given by~\ref{mass},~\ref{ade} and~\ref{darcy}, we have to assign the equations of state for $\mu_c({\bf x},t)$ and $\rho_c({\bf x},t)$. Here, linear constitutive relationships are adopted, namely
\begin{equation}
\rho_c({\bf x},t)=\rho_0 \left[1+\epsilon c({\bf x},t) \right]
\label{law1}
\end{equation}
and
\begin{equation}
\mu_c({\bf x},t)=\mu_0 \left[1+\gamma c({\bf x},t) \right],
\label{law2}
\end{equation}
which in most cases have been shown to accurately reproduce experimentally measured variations (see, e.g., the discussions in~\citep{gelhar1, tartakovsky_jfm, rev2}). Other functional forms are also possible (see, e.g.,~\citep{hassanizadeh, gelhar1, rev2} and references therein). The values $\mu_0$ and $\rho_0$ represent the reference viscosity and density of the fluid, respectively (when no contaminant is present). The parameters $\epsilon$ and $\gamma$ depend on the specific injected contaminant species and are defined as $\epsilon = (\rho_0^{-1}) \partial \rho_c/\partial c$ and $\gamma = (\mu_0^{-1})\partial \mu_c/\partial c$, evaluated at given experimental conditions. The magnitude of these parameters ultimately determines the strength of the nonlinear coupling between advection and concentration in Eq.~\ref{ade}.

Darcy's law \ref{darcy} can be conveniently rewritten in terms of the piezometric head $h=p/g \rho_0+y$, $y$ being the vertical axis coordinate:
\begin{equation}
{\bf u}({\bf x},t) = - \frac{K}{1+\gamma c({\bf x},t)} \left[ \nabla h + \epsilon c({\bf x},t){\bf e_y} \right],
\label{darcy2}
\end{equation}
where $K=k \rho_0 g / \mu_0$ is the hydraulic conductivity of the porous medium~\citep{bear} and ${\bf e_y}$ is the unit vector in the vertical direction. Then, combining Eqs.~\ref{darcy2} and~\ref{mass}, we finally have
\begin{equation}
\nabla \cdot \left \{ \frac{1}{1+\gamma c({\bf x},t)} \left[\nabla h + \epsilon c({\bf x},t){\bf e_y} \right] \right \} = 0
\label{piezo}
\end{equation}
i.e., a single equation for the unknown piezometric head $h$. In the resulting physical system provided by Eqs.~\ref{piezo},~\ref{ade} and~\ref{darcy2}, the unknowns to be determined are the piezometric head $h$, the advection field ${\bf u}({\bf x},t)$ and the concentration field $c({\bf x},t)$. Some authors mention that the variations of viscosity are less relevant than those of density, for typical transport problems in porous media (see, e.g.,~\citep{gelhar1, gelhar2}): we choose then to set $\mu_c \simeq \mu_0$ in the following. This hypothesis is to be tested {\em a posteriori} by comparing model outcomes with experimental measures.

\begin{figure}[t]
\centerline{\epsfxsize=9.0cm\epsfbox{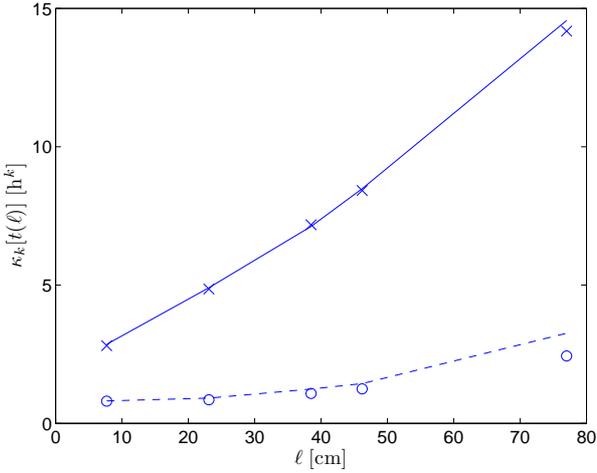}}
\caption{Downwards injection at a reference molarity $C^{mol}=0.2$ mol/L (cf.~Fig.~\ref{fig7}). Cumulants $\kappa_k [t(\ell)]$, $k=1,2$, of passage times $t(\ell)$ at various column heights $\ell$. Crosses represent the mean of the passage times ($k=1$), circles the variance ($k=2$) computed from experimental data. Solid ($k=1$) and dashed lines ($k=2$) are the model estimates.}
   \label{fig9}
\end{figure}

\begin{figure}[t]
\centerline{\epsfxsize=9.0cm\epsfbox{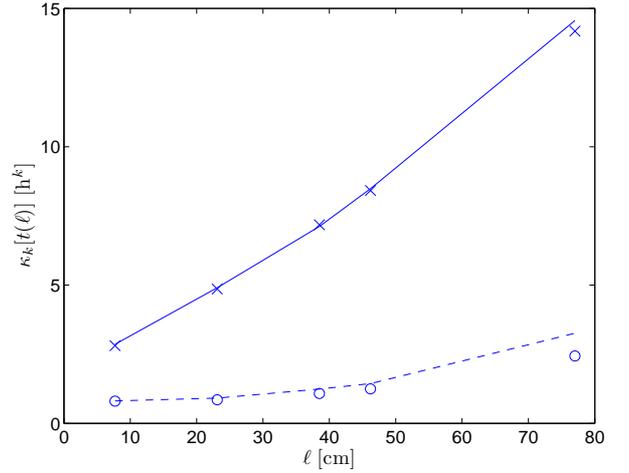}}
\caption{Upwards injection at a reference molarity $C^{mol}=0.2$ mol/L (cf.~Fig.~\ref{fig8}). Cumulants $\kappa_k [t(\ell)]$, $k=1,2$, of passage times $t(\ell)$ at various column heights $\ell$. Crosses represent the mean of the passage times ($k=1$), circles the variance ($k=2$) computed from experimental data. Solid ($k=1$) and dashed lines ($k=2$) are the model estimates.}
   \label{fig10}
\end{figure}

The system above is complemented by assigning the proper boundary and initial conditions, in agreement with the experimental configuration. On the lateral sides of the column we have ${\bf u} \cdot {\bf n} =0$ and $D \nabla c \cdot {\bf n} =0$, ${\bf n}$ being the normal outwards vector. This means that there is no flux of matter on the lateral sides. Note that this condition also implies $\nabla h \cdot {\bf n} =0$ on the lateral sides. Moreover, we impose a fixed value of the piezometric head $h$: it is expedient to set $h=0$ at the outlet of the column. As an initial condition, we assume that the concentration field is zero everywhere, $c({\bf x},0)=0$, as the porous material is saturated with the reference fluid. At the column inlet, we assign the ingoing flux in the form of a Robin (mixed) boundary condition:
\begin{equation}
-\left[{\bf u}({\bf x},t) - \theta D({\bf x},t) \nabla \right]c({\bf x},t) \cdot {\bf n} \vert_{\partial \Omega_{in} }  = J_0(t),
\end{equation}
where we have denoted by $\partial \Omega_{in}$ the entrance boundary. For the experimental runs that we have performed, the imposed flux $J_0(t)$ is modulated in time as a finite-duration step injection, from $t=0$ to $T$. Finally, we specify a Neumann boundary condition, $D \nabla c \cdot {\bf n} =0$, at the outlet of the column.

\section{Numerical solution}
\label{numerical}

Several alternative approaches have been proposed for solving the system of equations~\ref{piezo},~\ref{ade} and~\ref{darcy2}. In some special cases, it is possible to find approximate analytical solutions~\citep{oltean}, especially for stationary regimes~\citep{tartakovsky_jfm}. Aside, many efforts have been also devoted to the development of effective $1d$ models~\citep{tardy, hassanizadeh, landman1, landman2, landman3, liu}. These models are helpful when dealing with column devices, which can be regarded as being almost-$1d$. In general, however, one must resort to numerical solutions; for instance, Eulerian methods on a discretized domain, such as finite differences, finite elements or spectral methods, either applied to the model above, or to a set of ancillary equations for the flow lines (stream functions)~\citep{gelhar1, gelhar2, rogerson, ruith1, ruith2}. Fully Lagrangian methods are based instead on following fluid parcels along their trajectories through the porous medium~\citep{zoia}. As trajectories are correlated via the flow-transport coupling, this approach becomes computationally expensive for realistic multi-dimensional domains. Another Lagrangian approach, Smoothed Particles Hydrodynamics (SPH), has been recently proposed~\citep{sph, tartakovsky_prl}, which has shown good potential in dealing with Rayleigh-Taylor instabilities in porous media, especially when stochasticity is added to describe dispersion~\citep{tartakovsky_prl}. An hybrid approach, combining Eulerian and Lagrangian methods, consists in solving the flow equations by means of, e.g., finite differences on a grid, and then addressing transport via Monte Carlo simulation of contaminant particles stochastically moving on the grid, at assigned advection field~\citep{tchelepi, araktingi}. At each time step, the concentration profile derived by particle tracking is given as input to the numerical solver, in order to update the pressure distribution and re-compute advection. Finally, an increasing popular computational tool to tackle unstable interfacial dynamics is the Lattice Boltzmann Method: see, e.g.,~\citep{lbm} and references therein.

\begin{figure}[t]
\centerline{\epsfxsize=9.0cm\epsfbox{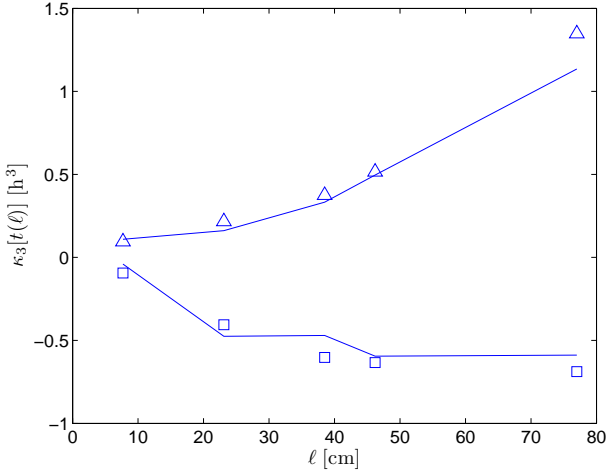}}
\caption{Central moment $\kappa_3 [t(\ell)]$ of passage times $t(\ell)$ at various column heights $\ell$, for downwards (squares) and upwards (triangles) injection at a reference molarity $C^{mol}=0.2$ mol/L. Model estimates for the two cases are drawn as solid lines.}
   \label{fig11}
\end{figure}

\begin{figure}[t]
\centerline{\epsfxsize=9.0cm\epsfbox{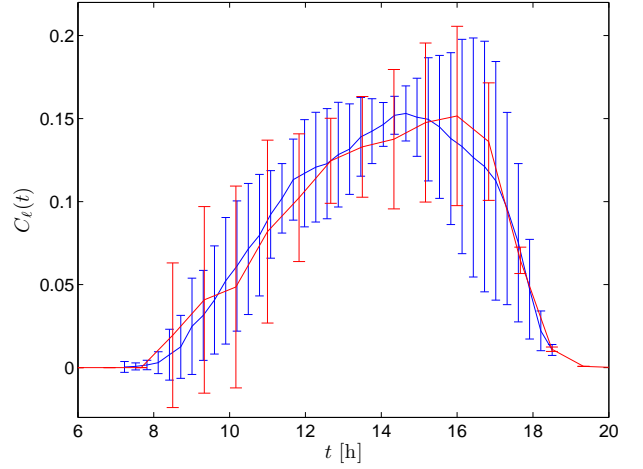}}
\caption{Downwards injection at a reference molarity $C^{mol}=0.5$ mol/L. Contaminant profile $C_\ell(t)$ measured at section $\ell =77$ cm, as a function of time [h]. Experimental data (averaged over 5 runs) are in blue, model outcomes (averaged over 5 initial random flux perturbations) are in red. Error bars for both curves correspond to one standard deviation.}
   \label{fig12}
\end{figure}

For our aims, we resort to a Eulerian approach and adopt the finite elements general-purpose code CAST3M (which has been developed at CEA~\footnote{CAST3M website: http://www-cast3m.cea.fr/cast3m/index.jsp.}) for solving the system~\ref{piezo},~\ref{ade} and~\ref{darcy2} above.

First, the time discretization of the equations is performed via a finite differences approach, by introducing a (small) parameter $\Delta t$ as the time step for integration. For sake of simplicity, we will consider the backward Euler implicit formula (BDF1); the extension to second-order accurate backward difference formula (BDF2) would be straightforward. In order to cope with the nonlinearities of the system above, we use a fixed-point approach. We suppose that, at a given discretized time $t^n=n \Delta t$, we know the discretized variables $h^n=h({\bf x},t^n)$, ${c}^n=c({\bf x},t^n)$, and ${{\bf u}}^n={\bf u}({\bf x},t^n)$. Let $j$ denote the index referring to the fixed-point iteration. Remark that, for $j=0$, the generic variable $a^{n+1,j}$ is equal to $a^{n}$. At the $j$-th fixed-point iteration, we first compute the discretized piezometric head $h^{n+1,j+1}$ by solving Eq.~\ref{piezo}, i.e.,
\begin{equation} 
\nabla \cdot \left[ \frac{1}{1+\gamma c^{n+1,j}} \left(\nabla h^{n+1,j+1} + \epsilon c^{n+1,j}{\bf e_y} \right) \right] = 0,
\label{h_disc}
\end{equation}
with the boundary conditions prescribed above.

Then, using Eq.~\ref{darcy2}, we compute the discretized advection field ${{\bf u}}^{n+1,j + 1}$ as follows
\begin{equation} 
{\bf u}^{n+1,j + 1} = - \frac{K}{1+\gamma c^{n+1,j}} \left( \nabla h^{n+1,j + 1} + \epsilon c^{n+1,j}{\bf e_y} \right).
\end{equation}
Finally, we compute the discretized concentration field $c^{n+1,j+1}$ using the discretized ADE~\ref{ade}:
\begin{equation}
\frac{c^{n+1,j+1} - c^{n+1,j}}{\Delta t} =\nabla \cdot \left[\theta D^{n+1,j+1} \nabla - {{\bf u}^{n+1,j+1}} \right]c^{n+1,j+1},
\label{ade_disc}
\end{equation}
with the boundary conditions above and
\begin{equation}
D^{n+1,j+1} = \alpha |{\bf u}^{n+1,j+1}|/\theta + D_0.
\end{equation}
Note that we have written the discretized equations including also viscosity variations and molecular diffusion, for sake of completeness. In actual calculations, these two terms are dropped. The convergence of the fixed-point iterations loop is achieved when the error on the generic variable, $|a^{n+1,j+1}-a^{n+1,j}|$, is below a given threshold. Then, the following time iteration is started, $t^{n+1}=t^n+\Delta t$.

The space discretization of the equations is performed via the finite element solver available in CAST3M. Within this code, several finite element types can be selected to discretize the variables and the space-dependent coefficients. In this work, both variables and coefficients are approximated by using Lagrange complete quadratic polynomials (Q2)~\citep{dhatt81}. These polynomials involve linear combinations of $1$, $x$, $y$, $x^2$, $xy$, $y^2$, $x y^2$, $x^2 y$, $x^2 y^2$ terms, in the frame of the reference element~\citep{dhatt81}. For our calculations, we typically used $10$ elements along the horizontal direction and $160$ along the vertical direction. This ratio was established on the basis of the aspect ratio of the column, i.e., $H/2r=16$.

Finally, the solution of the linear systems arising from the Poisson equation for $h$ (Eq.~\ref{h_disc}) and the discretised ADE equation for $c$ (Eq.~\ref{ade_disc}) is obtained via the Stabilized Bi-Conjugate Gradient method coupled with an ILUTP preconditioner, whose dimension is taken $1.5$ times larger than the matrix involved in the linear system~\citep{saad1, saad2}.

The time step $\Delta t$ is in principle arbitrary, and must be chosen small enough to attain convergence. In our numerical tests, it turns out that this condition is typically ensured when $\Delta t \le \min \left \{ s \Delta x , s \Delta y \right \}$, where $\Delta x$ and $\Delta y$ are the spacing of the finite elements in horizontal and vertical directions, respectively, and $s \simeq 0.02$. Tests of convergence were performed and gave satisfactory results.

\section{Testing model on experimental data}
\label{test}

In the following, we address the comparison between the experimental results described in Sec.~\ref{experiments} and the numerical simulations of the model, as exposed in Sec.~\ref{modeling} and~\ref{numerical}. The measured hydraulic conductivity is $K \simeq 1.59 \cdot 10^{-5}$ m/s, with $\rho_0=998.3$ Kg/m$^3$ and $\mu_0 = 0.997 \cdot 10^{-3}$ Kg/ms, which allows estimating the intrinsic permeability $k \simeq 1.62 \cdot 10^{-12}$ m$^2$. From the data in Table~\ref{tab1}, we have estimated $\epsilon \simeq (\Delta \rho/\rho_0)/\Delta c = 0.12$. Uncertainties on experimental values have been attributed to the parameters $\epsilon$ and $\alpha$, which have been adjusted so to improve the fit of model outcomes with respect to measured data.

Since our emphasis is on the spatial and temporal features of the contaminant plume dynamics, we commence by considering the evolution of concentration profiles measured at several heights along the column, as a function of time. In Fig.~\ref{fig7}, we display as a representative case the results for downwards injection at $C^{mol}=0.2$ mol/L. As experimental data correspond to section-averaged X-rays measures, simulated profiles have been as well obtained by averaging the ($2d$) concentration field $c({\bf x}, t) $ over the transverse direction $x$. To this aim, after computing the field $c({\bf x}, t)$ from the model, we define the section-averaged normalized concentration $C(y,t)=\int c({\bf x}, t) dx/B$ to make comparisons with experimental data possible.

As a general remark, the specific shapes of the unstable fronts measured in experiments (where the heavy fluid overlies the lighter one~\citep{wooding, taylor}) are not expected to be exactly reproduced by transverse-averaged simulated profiles; indeed, the observed profiles intrinsically depend on the random small-scale perturbations (such as different grain sizes, or preferential flow streams) encountered by the plume at injection and all along the column. These perturbations, which are ubiquitous even in (macroscopically) homogeneous media, vary at each experimental run and are ultimately responsible for the extension of the unstable front.

\begin{figure}
\centering
\subfigure[]{\includegraphics[width=.4\textwidth, angle=90]{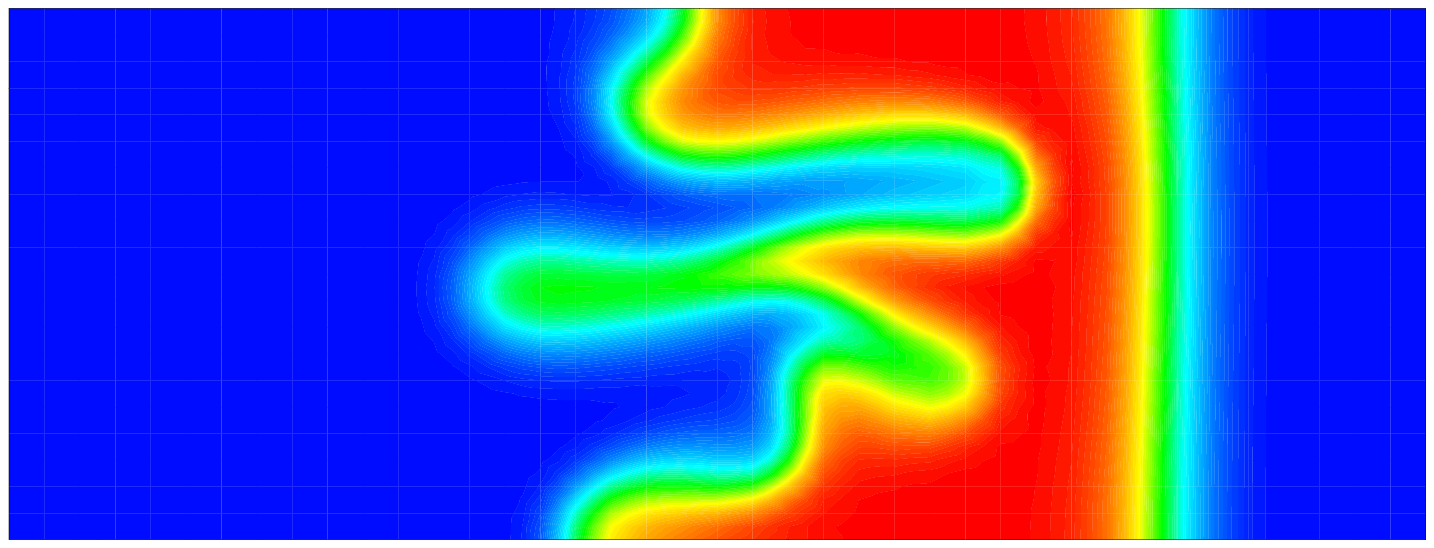}}
\hspace{.3in}
\subfigure[]{\includegraphics[width=.4\textwidth, angle=90]{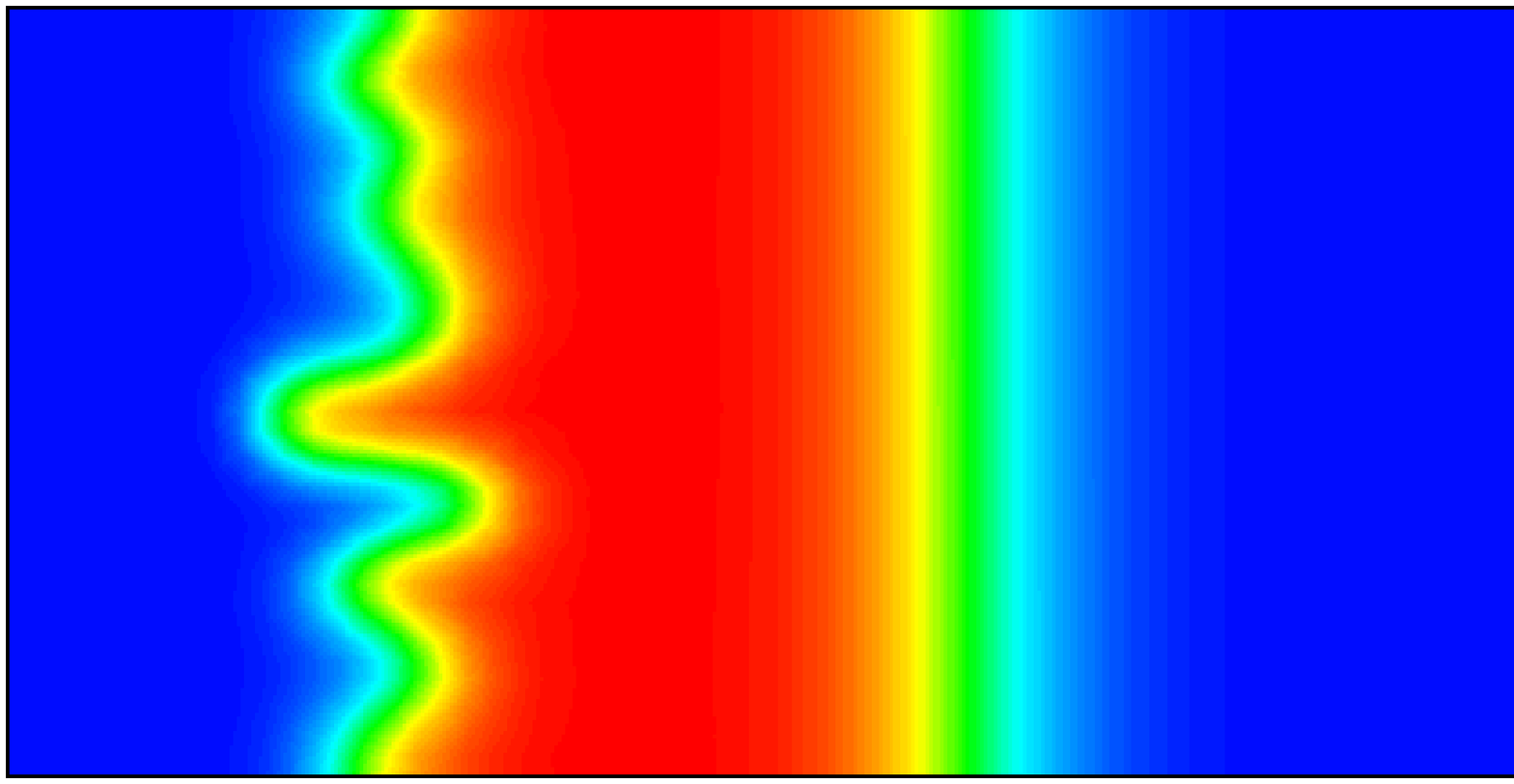}}
\caption{Concentration fields corresponding to downwards (left) and upwards (right) contaminant injection, at a given time. Stable and unstable fronts appear, depending on the density contrast between the injected (displacing) and resident (displaced) fluids. Color scale ranges from $C/C^{mol}=1$ (red) to $C/C^{mol}=0$ (blue).}
\label{fig13}
\end{figure}

\begin{figure}
\centering
\subfigure[]{\includegraphics[width=.4\textwidth, angle=90]{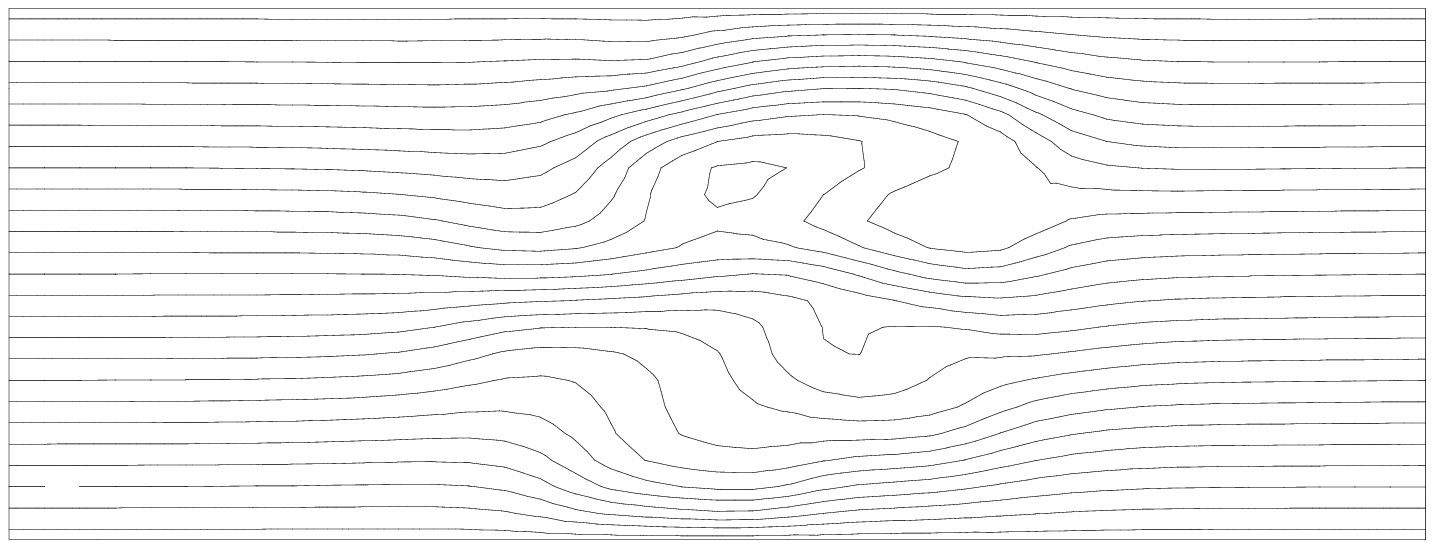}}
\hspace{.3in}
\subfigure[]{\includegraphics[width=.4\textwidth, angle=90]{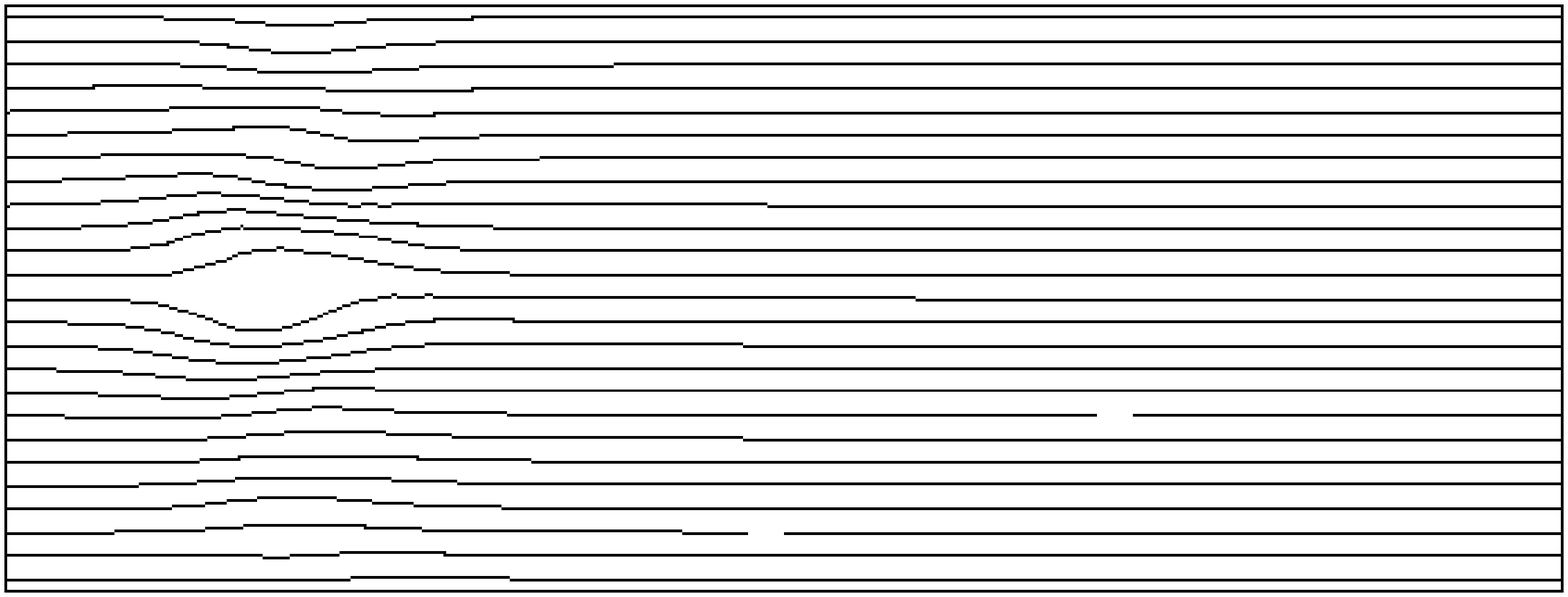}}
\caption{Steam lines corresponding to the cases presented in Fig.~\ref{fig13}, i.e., downwards (left) and upwards (right) contaminant injection, at a given time. Stream lines are distorted due to the coupling between advection and concentration fields. Distortion is evident where the contaminant  plume is located (cf.~Fig.~\ref{fig13}).}
\label{fig14}
\end{figure}

From the point of view of simulations, instabilities in the nonlinear system can be triggered by small random perturbations such as the unavoidable numerical noise that is produced by integrating the model equations~\citep{sph}. In practice, however, this procedure could require a time span that is longer than the simulation time, since the apparent geometric symmetries (the rectangular domain with a flat injecting surface) hinder the growth of such perturbations. Therefore, in order to force the appearance of interfacial instabilities, it is expedient to perturb the injected flux by superposing an additive white noise. In our computations, we have chosen a relative noise amplitude of $1 \%$: these small deviations are sufficient to trigger instabilities within the first simulation time steps. Physically, the additive noise could be justified on the basis of the small-scale fluctuations that are generated when injecting the contaminant inside the column: the adopted amplitude is well within instrument accuracy.

As discussed in Sec.~\ref{experiments}, the stronger the solution molarity $C^{mol}$, the stronger the effects of stochastic fluctuations on contaminant concentration profiles. However, the overall behavior at $C^{mol}=0.2$ mol/L, and especially the skewness of the curves, is well captured by the model at each position along the column. In particular, we may conclude that the approximations that we have introduced (such as considering a simplified geometry, neglecting molecular diffusion and viscosity variations) are {\em a posteriori} justified on the basis of the comparison between model and data. In Fig.~\ref{fig8}, we perform a similar analysis for upwards injection at $C^{mol}=0.2$ mol/L, for the same step-injection flow conditions. Also in this case, a good quantitative agreement is found between simulated and measured concentration profiles.

Up to a normalization factor, the transversally-averaged concentration $C(\ell,t)dt$ can be interpreted as the distribution function of the contaminant plume passage times $t(\ell)$ at various heights $y=\ell$ along the column. Then, the comparison between model and experimental data is substantiated in Figs.~\ref{fig9} and~\ref{fig10} by computing the first two cumulants of the passage times, i.e.,
\begin{eqnarray}
\kappa_1 [t(\ell)]=\int t C(\ell,t) dt \nonumber \\
\kappa_2 [t(\ell)]=\int (t-\kappa_1)^2 C(\ell,t) dt,
\end{eqnarray}
for the cases of Figs.~\ref{fig7} and~\ref{fig8}, respectively. The quantity $\kappa_1 [t(\ell)]$ [h] is the mean of the distribution and defines the average time required by the contaminant plume to reach height $\ell$ from the injection position; the quantity $\kappa_2 [t(\ell)]$ [h$^2$] is the variance and defines the (squared) spread around the average time. Model outcomes provide accurate estimates of the moments computed from experimental data. Note in particular that the quantity $v = \ell / \kappa_1 [t(\ell)]$ identifies the average velocity of the pollutants in the porous media: to a first approximation, it turns out that $v \simeq v_0$, where $v_0 \simeq 6$ cm/h is the nominal plume velocity in absence of flow-transport couplings. This implies that the average velocity is minimally affected by density variations (for the experimental conditions considered here). On the contrary, fluctuations around the mean velocity do not simply average out, and are responsible for the increased dispersion in correspondence of the unstable front between resident and displacing fluids, i.e., the skewed concentration profiles.

The deviation from Gaussianity of the concentration profiles shown in Figs.~\ref{fig7} and~\ref{fig8} is better elucidated by means of the third cumulant $\kappa_3 [t(\ell)]=\int (t-\kappa_1)^3 C(\ell,t) dt$ [h$^3$], which is displayed in Fig.~\ref{fig11}. For Fickian transport, Gaussian (symmetrical) concentration profiles would lead to vanishing $\kappa_3 [t(\ell)]$; on the contrary, the moment estimates computed from experimental data indicate that downwards injection leads to negatively skewed profiles ($\kappa_3 [t(\ell)]<0$), whereas upwards injection leads to positively skewed profiles ($\kappa_3 [t(\ell)]>0$). Remark in particular that the absolute value of the third moment is an increasing function of column length $\ell$, which means that deviations from Gaussianity become more apparent as the contaminant plume moves through the porous material. Model estimates, which are displayed in the same figure, are in good agreement with experimental data.

Finally, in order to further support our analysis, we analyze the concentration profiles at the last measure point along the column ($\ell =77$ cm) at $C^{mol}=0.5$ mol/L, for downwards injection at the same flow conditions as above. We know from previous discussions that concentration profiles at this molarity show apparent fluctuations that hinder reproducibility; model simulations show a similar behavior. Then, in order to make comparisons possible, we choose to average measured concentration curves over several experimental runs, and to average model outcomes over several random realizations of the initial flux perturbations. In Fig.~\ref{fig12}, we display the averaged profiles as well as the error bars (corresponding to one standard deviation). Although the model can not capture the precise shape of the single concentration profiles, a remarkable quantitative agreement is found for the averaged realizations. In other words, the model is actually capable of correctly predicting the average displacement and the spatial extension (spread) of the pollutant plume, even at strong molarities. Similar results have been obtained also for upwards injection (not shown here).

\section{Qualitative features of variable-density transport}
\label{qualitative}

The finite elements numerical tool described in Sec.~\ref{numerical} provides a practical means of exploring the contaminant dynamics inside the column, and can also be used to extract informations that are not easily accessible (or even not available at all) from experiments. In particular, we are interested in exploring the general qualitative features of the interfacial behavior. In fact, the X-rays measures correspond to averaging the concentration profiles in the transverse dimension at any fixed height along the column, and thus hide the fine-scale properties of such phenomena as fingering and instabilities.

Typical simulated concentration fields $c(x,y)$ corresponding to variable-density contaminant transport are displayed in Fig.~\ref{fig13} at a fixed time $t=2.1$ h, for both downwards and upwards injection (with duration $T=1.25$ h). For this simulation we chose $C^{mol}=0.3$ mol/L, with column geometry $H=40$ and $B=15$. We remark that, because of the flux shape (with a finite time duration), two fronts appear at the interface between injected and resident fluids: one is unstable (owing to the density contrast with respect to the resident fluid) and displays fingers, whereas the other is stable. In correspondence of the stable front, a standard Gaussian concentration profile is found. The unstable front gives rise to an enhanced extension of the mixing (dispersive) region of the injected plume, in agreement with experimental observations~\citep{dangelo, debacq}. Note that the extension of the fingers is not entirely symmetric upon reversal of the flow direction.

We observe that, once triggered, unstable fingers can grow only provided that the contaminant molarity is sufficiently strong to overcome the opposing effects of dispersion, which acts as a smoothing process on the concentration profiles. When transport is Fickian (i.e., at weak molarity), our simulations show that dispersion dominates and the small flux perturbations are rapidly reabsorbed, so that after a few time steps the contaminant profiles become perfectly smooth. In particular, the shape of the concentration profiles becomes independent of the specific realization of the initial flux perturbation. This is indeed coherent with the experimental findings of, e.g.,~\citep{dewit}. The number of fingers that can actually appear and survive the dispersive smoothing strongly depends on the geometry of the column: in general larger diameters lead to an increased number of such fingers (as naturally expected), for a given value of the molarity~\citep{taylor}.

The advection field ${\bf u}({\bf x},t)$ is also deeply affected by density variations. For illustrative purposes, in Fig.~\ref{fig14} we plot the stream lines corresponding to the case presented Fig.~\ref{fig13}. The stream function $\psi(x,y)$ is defined as the scalar function whose contour lines are the stream lines. For two-dimensional flows satisfying Eq.~\ref{mass}, $\psi(x,y)$ is such that
\begin{eqnarray}
{\bf u}_x = \frac{\partial \psi(x,y)}{\partial y} \nonumber \\
{\bf u}_y = - \frac{\partial \psi(x,y)}{\partial x}
\label{stream}
\end{eqnarray}
at any fixed time $t$, ${\bf u}_x$ and ${\bf u}_y$ being the horizontal and vertical components of the velocity field ${\bf u}(x,y)$, respectively. Then, it follows that $\Delta \psi(x,y)=-\nabla \times {\bf u}(x,y)$. By definition, the stream lines are instantaneously tangent to the velocity vector. From Fig.~\ref{fig14}, it is immediately apparent that stream lines for both downwards and upwards injection are perturbed where the contaminant plume is located, by virtue of the coupling between concentration and advection. This is to be compared with the expected behavior of Fickian transport, where the vertical component of the velocity field is constant (and equal to $v_0$) and the transversal component vanishes by virtue of symmetries, so that the advection field is homogeneous and stream lines are parallel and oriented along the vertical direction.

Finally, observe that the time duration $T$ of injection can play an important role, as experimentally investigated, e.g., in~\citep{wood, dewit}. Indeed, while for two semi-infinite fluids only their mutual interface matters, in our case two distinct fronts arise. At the unstable front, dense fluid fingers move downwards under the influence of gravity, whereas light fluid fingers move upwards due to buoyancy forces (this is particularly evident in Fig.~\ref{fig13} for downwards injection). We have numerically verified that, if the time duration of the injection is short (so that the spatial extension of the contaminant plume is limited), light fluid fingers might reach the stable front at the opposite end and disrupt the integrity of the plume. At longer time scales, dispersion eventually takes over and the fingering dies out, because of dilution~\citep{dewit}. These phenomena have been tested numerically, but do not actually occur for the experimental conditions under investigation here.

\section{Conclusions}
\label{conclusions}

In this work, we have addressed the problem of variable-density transport of a miscible fluid, with the aim of ascertaining the spatial and temporal features of such contaminant dynamics in porous media. This topic has been the subject of intense research activity, by virtue of its relevance in such technological challenges as polluted site remediation, and enhanced oil recovery. Our investigation has been supported by novel experimental results.

Data have been collected by means of a vertical column setup filled with homogeneous saturated sand, coupled with a dichromatic X-ray source and associate detector. As contaminant spills in groundwater typically have a finite extent, our analysis has focused on finite-duration pollutants injections. Experimental results show that the concentration profiles are skewed because of gravity effects, and that the sign of the skewness depends on the flow direction.

In order to gain a deeper insight on the underlying physical phenomena, we have briefly reviewed the nonlinear equations that govern variable-density transport and determined their numerical solutions by resorting to the finite elements code CAST3M. We have then compared the simulation results with the experimental data: the model actually captures the salient features of the contaminant dynamics and displays a good quantitative agreement with measurements.

Finally, the finite elements model has been also used as a means of exploring the behavior of the interfacial dynamics between the dense contaminant plume and the lighter resident fluid that saturates the column, whose fine-scale properties are not accessible by experiments. Such interfacial dynamics is ultimately responsible for the skewed shape of the contaminant profiles within the column. For downwards injection, gravitational instabilities appear at the front of the plume, and the enhanced mixing is such that contaminant reaches the outlet earlier than expected when adopting the standard ADE approach. At the opposite, for upwards injection the instabilities appear at the rear of the plume, and contaminant reaches the outlet of the column later than expected.

In order to perform the comparison between model and data, some simplifying assumptions have been introduced, which surely deserve further consideration. For instance, it would be important to isolate the effects of viscosity (running tests on horizontal columns) and molecular diffusion (at very slow flow rates or with ad hoc devices such as in~\citep{kirino}). Moreover, we should separately test the relevance of the Boussinesq approximation and extend the numerical simulations to $3d$ cylindrical geometry.

Many issues remain to be addressed, in view of such applications of our laboratory-scale study as to polluted site remediation. A fundamental question is whether the present results can be extended to larger domains (field-scale measures) via simple scaling arguments, or more complex mechanisms and couplings must be taken into account. In particular, it has been pointed out that the validity of Darcy's and Fick's laws might be questionable~\citep{hassanizadeh}. As larger scales unavoidably involve complex spatial patterns, preliminary experimental work will concern the case of heterogeneous and/or non-saturated media, where deviations from Fickian behavior due to the non-homogeneous structure of the porous material may compete with those of density. In this context, preliminary considerations seem to suggest that a suitable modeling tool could be provided by an extension of the continuous time random walk formalism to nonlinear transport~\citep{zoia}.

\section*{Acknowledgments}
The authors would like to thank E.~Studer and S.~Gounand (LTMF) for support in developing the numerical model in CAST3M and Ph.~Montarnal for constructive comments.

\end{document}